\documentclass[reprint,aps,floats,floatfix,amsmath,amssymb,amsfonts,nofootinbib,longbibliography,superscriptaddress]{revtex4-1}

\usepackage{graphicx}
\usepackage{dcolumn}
\usepackage{bm}
\usepackage{tensor} 
\usepackage{natbib}
\usepackage{subfigure}

\usepackage[hyperindex,colorlinks]{hyperref}

\begin{document}

\title{Revisiting $f(R,T)$ cosmologies}

\author{Ana Paula Jeakel}
\email{ana.pj.dias@edu.ufes.br}
\affiliation{%
N\'ucleo Cosmo-ufes \& Departamento de F\'isica,  Universidade Federal do Esp\'irito Santo (UFES)\\
Av. Fernando Ferrari, 540, CEP 29.075-910, Vit\'oria, ES, Brazil.}%

\author{Jonas Pinheiro da Silva}%
\email{jonas.j.silva@edu.ufes.br}
\affiliation{%
PPGCosmo, CCE, Universidade Federal do Esp\'irito Santo (UFES)\\
Av. Fernando Ferrari, 540, CEP 29.075-910, Vit\'oria, ES, Brazil.}%

\author{Hermano Velten}%
\email{hermano.velten@ufop.edu.br}
\affiliation{%
Departamento de F\'isica, Universidade Federal de Ouro Preto (UFOP), Campus Universit\'ario Morro do Cruzeiro, 35.400-000, Ouro Preto, Brazil}%

\date{\today}

\begin{abstract}We review the status of $f(R,T)$ cosmological models, where $T$ is the trace of the energy momentum tensor $T^{\mu\nu}$. We start focusing on the modified Friedmann equations for the minimally coupled gravitational Lagrangian of the type $f(R,T)=R +\alpha e^{\beta T} + \gamma_{n} T^{n}$.  We show that in such a minimally coupled case there exists a useful constraining relation between the effective fractionary total matter density with an arbitrary equation of state parameter and the modified gravity parameters. With this association the modified gravity sector can be independently constrained using estimations of the gas mass fraction in galaxy clusters. Using cosmological background cosmic chronometers data and demanding the universe is old enough to accommodate the existence of Galactic globular clusters with ages of at least $\sim 14$ Gyrs we find a narrow range of the modified gravity free parameter space in which this class of theories remains viable for the late time cosmological evolution. This preferred parameter space region accommodates the $\Lambda$CDM limit of $f(R,T)$ models. We also work out the non-minimally coupled case in the metric-affine formalism and find that there are no viable cosmologies in the latter situation. However, when analysing the cosmological dynamics including a radiation component, we find that this energy density interacts with the matter field and it does not scale according to the typical behavior. We conclude stating that $f(R,T)$ gravity is not able to provide a full cosmological scenario and should be ruled out as a modified gravity alternative to the dark energy phenomena.
\end{abstract}

\maketitle
\onecolumngrid
\section{Introduction}

Dark matter (DM) and dark energy (DE) compose the so-called dark sector of the universe and represent intriguing elements of modern cosmology. Whereas the former is responsible for many unexpected astrophysical observations e.g., flatness of galaxy rotation curves, and also plays a crucial role in the cosmological large scale structure formation, the latter is evoked to deal with the current accelerated phase of the background expansion rate firstly denounced by Supernovae type Ia observations. Conversely, the inclusion of both components in the standard cosmological model can be understood as the inability of General Relativity to properly describe the gravitational interaction at scales beyond the Galactic one. This has motivated the rise of a new research route where one searches for extensions/modifications of the Einstein-Hilbert Lagrangian.

There are many distinct ways to go beyond General Relativity see Ref. \cite{Clifton:2011jh,Nojiri:2017ncd} for a review. Apart from adding new fields, departures from Riemannian geometries or adopting quantum arguments, perhaps, the most natural way to modify gravity is to add invariants in the Einstein-Hilbert Lagrangian giving rise to higher-order theories. The widely known prototype within this category is the set of $f(R)$ theories \cite{DeFelice:2010aj,Nojiri:2010wj}. In the latter, the Einstein-Hilbert Lagrangian, term $f_{\textsl{EH}}(R)=R$, where $R=g_{\mu\nu}R^{\mu\nu}$ is the Ricci scalar, $g_{\mu\nu}$ is the metric and $R^{\mu\nu}$ is the Ricci tensor, is replaced by a more general algebraic combination of $R$. By going beyond $f(R)$ theories, one can keep adding geometric invariants to the gravitational Lagrangian or, for instance, to implement a non-minimal coupling between geometry and matter fields. Within the latter strategy, two classes of theories have appeared recently, the $f(R, L_{m})$ gravity \cite{Harko:2010mv}, where $L_{m}$ is the matter Lagrangian and the $f(R,T)$ gravity \cite{Harko:2011kv}, where $T=g_{\mu\nu}T^{\mu\nu}$ is trace of the energy-momentum tensor.

In this work we will study $f(R,T)$ theories as an alternative to the dark energy phenomena with focus on their cosmological background expansion. Several $f(R,T)$ solutions for the cosmological expanding background have been found in the literature. Analytical reconstructions of dark energy phenomenological models associated to $f(R,T)$ gravity has been studied in Refs. \cite{Jamil:2011ptc,Houndjo:2011fb,Houndjo:2011tu,Odintsov:2013iba}. 
The asymptotic behavior of the scale factor and confrontation with data has been performed in e.g., Refs. \cite{Velten:2017hhf,Sahoo:2018rhn, Nagpal:2019vre,Shabani:2017lgy,Goncalves:2022ggq}. The thermodynamical implications of this theories have been studied in \cite{Jamil:2012pf}. The rich phenomenology of the universe's expansion can also be fully addressed by $f(R,T)$ theories using its scalar-tensor version \cite{Lobao:2023wde}. Most of the $f(R,T)$ models are capable to induce a late time accelerated expansion rate providing negative values for the today's deceleration parameter $q_0$. However, in light of available modern cosmological data, a truly viable model should obey several other requirements. 

Ref. \cite{Velten:2017hhf}, by one of the authors, has challenged some of the available $f(R,T)$ models by arguing that though the low-z evolution of $f(R,T)$ models can be reasonable supported by available data, there is a considerable discrepancy in the high-z ($z>1$) dynamics in comparison with standard $\Lambda$CDM cosmology. Then, this reference concludes that the viability of $f(R,T)$ cosmological models is severely challenged.   

Now, in this work, by considering a broad class of $f(R,T)$ cosmologies and using additional information about the age of the universe and the existing bounds on the gas mass fraction in galaxy clusters, we will revisit this issue. We shall consider that viable modified gravity based cosmologies $i)$ are able to reproduce quantitatively low redshift data as, for example, $H(z)$ data and, $ii)$ can yield to a minimum value for the universe's age consistent with the oldest astrophysical objects found so far and $iii)$ their modified gravity free parameters are constrained such that the effective fractionary matter density parameter is consistent with available  gas mass fraction data in galaxy clusters. Requirements $i$ and $ii$ are the new considered aspects in this present work in comparison with the analysis done in Ref. \cite{Velten:2017hhf}. The age argument is motivated since age estimations of globular clusters in our galaxy are available. Such estimations set a conservative lower bound of $t_U \gtrsim 14.16$ Gyrs for the age of the universe \cite{
Valcin:2021jcg,Valcin:2020vav}. Also, estimations from the gas mass fraction within galaxy clusters  obtained in \cite{Mantz:2014xba} place bounds on the cosmological fractionary total matter density parameter $\Omega_0$. 

In the next section we review  the cosmological background dynamics for $f(R,T)$ theories. The observational analysis is performed in section \ref{section3}.  We also investigate the inclusion of a radiation component in $f(R,T)$ cosmologies in section \ref{section5}. As an additional analysis, the non-minimally coupled case is studied in section \ref{section4}. We conclude in the final section.

\section{Cosmological background expansion in $f(R,T)$ theories}\label{section2}
 
The total action $S$ for the $f(R,T)$ theories \textbf{reads \cite{Harko:2011kv}}
\begin{equation}\label{action}
    S = \frac{1}{2\kappa ^{2}}\int \sqrt{-g}\,d^{4}x\,f(R, T) + \int \sqrt{-g}\,d^{ 4}x\,L_{m}(g_{\mu\nu}, \psi _{m}),
\end{equation}
where $\kappa ^{2} = 8\pi G$ is the coupling constant, $g$ is the metric determinant and $L_{m}$ is the Lagrangian for the matter sector gathering the contribution of all matter fields $\psi_m$. In this work,  since we are focused on the late time cosmological dynamics, we consider that $L_m$ is composed by a perfect fluid representing the total (dark + baryonic) matter contribution.   

By applying the variational principle to the above action \textbf{one finds \cite{Harko:2011kv}}
\begin{equation}\label{eq}
    f_{R}R_{\mu\nu} - \tfrac{f(R, T)}{2}g_{\mu\nu} - \Delta _{\mu\nu}f_{R} = \kappa ^{2}T_{\mu\nu}\left(1 - \frac{f_{T}}{\kappa ^{2}}\right) - f_{T}\Theta _ {\mu\nu}.
\end{equation}
In the equation above, we have used the notation 
\begin{equation}
    f_{R} \equiv \frac{\partial f(R, T)}{\partial R} \quad {\rm and}\quad f_{T} \equiv \frac{\partial f(R, T)}{ \partial T}.
\end{equation}
It is worth noting that the variation of the Ricci tensor has an explicit dependence \textbf{on the metric \cite{Harko:2010mv}}
\begin{equation}
    g^{\mu\nu}\delta R_{\mu\nu} = -\Delta _{\mu\nu}\delta g^{\mu\nu},
\end{equation}
where the d'Alembertian is related to the Ricci tensor by $\Box = g^{\beta \alpha}\nabla _{\beta}\nabla _{\alpha}$ and $\Delta _{\beta \alpha} = \nabla _{\beta}\nabla _{\alpha} - g_{\beta \alpha}\Box$. 

The quantity $\Theta_{\mu\nu}$ has been introduced according to the following reasoning. In order to characterise the matter sector, the energy-momentum tensor is defined as usually by 
\begin{equation}\label{em}
    T_{\mu\nu} = - \frac{2}{\sqrt{-g}}\frac{\delta \left(\sqrt{-g}L_{m}\right)}{\delta g^{\mu\nu}}.
\end{equation}
Then, the variation of this quantity can be written as
\begin{equation}
    \delta T = \left(\Theta _{\mu\nu} + T_{\mu\nu}\right)\delta g^{\mu\nu},
\end{equation}
where the auxiliary quantity $\Theta_{\mu\nu}$ appearing in \eqref{eq} has been defined as
\begin{equation}\label{eqtheta}
\Theta _{\mu\nu} \equiv g^{\alpha \beta}\frac{\delta T_{\alpha \beta}}{\delta g^{\mu\nu}} = - 2T_{\mu\nu} + g_{\mu\nu}L_{m} - 2g^{\alpha \beta}\frac{\partial ^{2}L_{m}}{\partial g^{\alpha \beta}\partial g^{\mu\nu}}.
\end{equation}

For pedagogical purposes, by reintroducing \eqref{eqtheta} into \eqref{eq}, it is also convenient to rewrite the field equation in the following way,
\begin{equation}\label{eq8}
    f_R R_{\mu\nu} - \tfrac{f(R, T)}{2}g_{\mu\nu} - \Delta _{\mu\nu}f_{R} = (\kappa ^{2} + f_{T})T_
{\mu\nu} + f_{T}\vartheta _{\mu\nu},
\end{equation}
where, 
\begin{equation}
    \vartheta _{\mu\nu} \equiv 2g^{\alpha \beta}\frac{\partial ^{2}L_{m}}{\partial g^{\alpha \beta}\partial g^{\mu\nu}} - g_{\mu\nu}L_{m}.
\end{equation}
Equation \eqref{eq8} shows explicitly the effective coupling term $\kappa^2 + f_T$ to the energy-momentum tensor.

By adopting the perfect fluid structure for the energy momentum tensor one reads
\begin{equation}
    T_{\mu\nu} = \left(\rho + p\right)u_{\mu}u_{\nu} - pg_{\mu\nu},
\end{equation}
with the four-velocity in comoving coordinates $u_{\nu} = (1, 0, 0, 0)$ and $\rho$ and $p$ being the energy density and pressure, respectively. 

Let us then apply to this set of equations a flat, homogeneous, isotropic and expanding spacetime given by the Friedmann-Lemaître-Robertson-Walker metric (FLRW)
\begin{equation}\label{flrw}
    ds^{2} = dt^{2} - a(t)^{2}\left(dr^{2} + r^{2}d\theta ^{2} + r^{2}\sin^{2}\theta d\phi ^{2}\right),
\end{equation}
where $a(t)$ is the cosmological scale factor and we consider $c^2=1$.

The $0-0$ component of $f(R,T)$ gravity \eqref{eq8} will provide an expression for the expansion rate $H=\dot{a}/a$.
The dot means a derivative with respect to the cosmic time. Since the Ricci scalar reads $R=-6(\ddot{a}/a + H^2)$ the modified Friedmann equation in $f(R,T)$ cosmology becomes
\begin{eqnarray}
    &H^{2}& = \frac{\kappa^{2}}{3}\left\{\rho  - \frac{\rho(1 + f_{R})}{f_{R}}- \frac{\left[f_{T}\rho (1 + \omega) + f(R,T)/2 -3H\dot{f}_{R} + 3\dot{H}f_{R}\right]}{\kappa ^{2}f_{R}}\right\}
\end{eqnarray}
The above equation has the appropriate GR limit with $f (R,T)=R, f_R=1, f_T=0$\textbf{ and also the} $f(R)$ gravity limit of $f_T=0$. Also, $\omega$ is the equation of state parameter relating the energy density $\rho$ to the pressure by $\omega=p/\rho$.

The expansion rate can then be rewritten in a compact form as
\begin{equation}\label{eqf}
    3H^{2} = \kappa ^{2}\left(\rho + \Bar{\rho}\right),
\end{equation}
where
\begin{eqnarray}\label{rhobar}
   \bar{\rho} &=& - \frac{1}{\kappa ^{2}f_{R}}\left[\kappa ^{2}\rho(1 + f_{R}) + f_{T}\rho (1 + \omega) + \tfrac{f(R,T)}{2} -3H\dot{f}_{R} + 3\dot{H}f_{R}\right].
\end{eqnarray}

The quantity $\rho$ should be interpreted as the sum of all matter fields composing the total energy momentum tensor of the theory. In the standard cosmology it can be approximated by the sum of radiation, matter (dark + baryonic) and a dark energy component. The modified gravity contribution to the expansion rate can be collected in terms of the geometrical effective energy density $\bar{\rho}$. This can be associated with the dark energy sector but here written in terms of the geometrical quantities and $\rho$ as well. If the modified gravity sector is responsible for the late time accelerated phase, then $\rho$ can be approximated, at late times, by the total matter. This is the interpretation we adopt in this work.

The complete description of the cosmological background expansion demands the second Friedmann equation obtained with the spatial components of \eqref{eq8}. It reads
\begin{equation}\label{eqf2}
(\dot{H} + 3H^{2})f_{R} + \tfrac{f(R, T)}{2} -2H\dot{f}_{R} - \ddot{f}_{R} = \kappa ^{2} p.
\end{equation}

It is worth noting that $f(R,T)$ theories are non-conservative since they present a non-vanishing covariant derivative of the energy momentum tensor as given by the expression
\begin{equation}\label{eqf4}
    \dot{\rho} + 3H\rho(1 + \omega) = - \frac{1}{\kappa ^{2} + f_{T}}\left[\dot{f}_{T}\rho(1 + \omega) + f_{T}\dot{\rho}\omega + \frac{\dot{f}(T)}{2}\right].
\end{equation}

The above equations apply for any $f(R,T)$ model. Only in a few cases the chosen $f(R,T)$ function leads to a vanishing right hand side of \eqref{eqf4}. Apart from this specific case, $f(R,T)$ cosmological models are non-conservative and the effective matter density parameter will no longer scale as $\rho \sim a^{-3}$. For a complete discussion on the issue of conservation of the energy momentum tensor in $f(R,T)$ theories see \cite{Bertini:2023pmp}.

In order to go further one has to specify the functional form of $f(R,T)$. The simplest assumption is the minimally coupled case in which the contributions from $R$ and the trace $T$ are written separately as
\begin{equation}
    f(R,T)= f_1(R)+f_2(T).
\end{equation}
Keeping this format the most general function covering the main proposals in the literature can be written as
\begin{equation}\label{model}
    f(R, T) = R + \alpha e^{\beta T} + \gamma _{n}T^{n}.
\end{equation}
This model has four free parameters $\alpha, \beta, \gamma_n$ and $n$. All power law models proposed in the literature are reached with $\alpha=0$. Also, the recently proposed exponential model (see. Ref. \cite{Moraes:2019hgx}) is equivalent to $\gamma_n=0$. General Relativity with no cosmological constant (the Einstein-de Sitter universe) corresponds to $\alpha=\gamma_n=0$. For $\beta = 0$ and $n = 0$ the $\Lambda$CDM model is recovered.

It is convenient to rewrite the background equations replacing $\rho$ by the fractionary density $\Omega=\rho/\rho_{c0}$, where $\rho_{c0}$ is the today's critical density $\rho_0 = 3H^{2} _{0}/\kappa ^{2}$. Then, according to \eqref{model} the FLRW expansion rate in  $f(R,T)$ theories reads 
\begin{eqnarray}\label{Hsquare}
    &&\frac{H^{2}}{H_{0}^{2}} = \Omega + \Bar{\alpha}e^{\Bar{\beta}\Omega(1 - 3\omega)}\left[\Bar{\beta}\Omega (1 + \omega) + \frac{1}{2}\right] + \Bar{\gamma} _{n}\left[n(1 + \omega)(1 - 3\omega)^{n - 1} + 
    \frac{1}{2}(1 - 3\omega)^{n}\right]\Omega ^{n}.
\end{eqnarray}
In the above expression we have rewritten the modified gravity free parameters in a dimensionless form according to
\begin{equation}
    \Bar{\alpha} = \frac{\alpha}{\kappa ^{2}\rho _{0}}; \hspace{0.5cm} \Bar{\beta} = \beta \rho _{0} \hspace{0.5cm} \Bar{\gamma}_{n} = \frac{\gamma_{n} \rho _{0}^{n - 1}}{\kappa ^{2}}.
\end{equation}
The cosmological dynamics will be obtained as a function of the fractionary density parameter $\Omega$. This quantity is obtained by rewriting \eqref{eqf4} in terms of dimensionless quantities defined above such that

\begin{eqnarray}\label{Omegadot}
   \dot{\Omega} + 3H\Omega (1 + \omega) &&= - \frac{\dot{\Omega}}{1 + \Bar{\alpha}e^{\Bar{\beta}\Omega(1 - 3\omega)}\Bar{\beta} + \Bar{\gamma}_{n}n\Omega ^{n - 1}(1 - 3\omega)^{n -1}} 
     \times\biggl\{\Bar{\alpha}e^{\Bar{\beta}\Omega(1 - 3\omega)}\Bar{\beta}\left[\Bar{\beta}\Omega (1 + \omega) + \omega + \frac{1}{2}\right] + \nonumber\\
     &&+\Bar{\gamma}_{n}n\Omega ^{n - 1}(1 - 3\omega)^{n -1}\left[\frac{2n(1 + \omega) - (1 + 3\omega)}{2}\right]\biggr\}.
\end{eqnarray}
 
The numerical solution of the set of equations presented above will allow us to analyze the background expansion in $f(R, T)$ theories. The first step for solving it is to set today's value $\Omega(z=0)\equiv\Omega_0$ as the initial condition for this differential equation. Once more, this quantity is interpreted as the total (baryonic + dark) matter fraction. This quantity is not a free parameter since it is subjected to the constraining relation
\begin{eqnarray}\label{Omega0}
    1 &=& \Omega_0 + \Bar{\alpha}e^{\Bar{\beta}\Omega_0(1 - 3\omega)}\left[\Bar{\beta}\Omega_0 (1 + \omega) + \frac{1}{2}\right] + \Bar{\gamma} _{n}\left[n(1 + \omega)(1 - 3\omega)^{n - 1} + 
    \frac{1}{2}(1 - 3\omega)^{n}\right]\Omega_0 ^{n}.
\end{eqnarray}
This relation appears from \eqref{Hsquare} by setting $H(z=0)=H_0$. Therefore, today's effective fractionary matter parameter $\Omega_{0}$ can not be arbitrarily chosen. This is a very important aspect we want to highlight since, with the exception of Ref. \cite{Velten:2017hhf}, this is not usually considered in previous analysis of the background expansion in $f(R,T)$ theories. This is possible since the adopted $f(R,T)$ function is minimally coupled. This means that $H^2$ does not depend on $\dot{H}$. As discussed further, in the non-minimally coupled case one can not obtain a similar constraining relation as well as in the case of $f(R)$ theories. In both cases, using the metric formalism,  $H^2$ depends on $\dot{H}$ and a constraining relation like \eqref{Omega0} does not exist. By switching off the modified gravity contributions with $\bar{\alpha}=\bar{\gamma}_n=0$ one recovers the Einstein-de Sitter model $\Omega_0=1$. For non-vanishing $\bar{\alpha}$ and $\bar{\gamma}_n$ values, and demanding $0<\Omega_0<1$ it is possible to place bounds on the possible values for the modified gravity parameters. 


\section{Observational constraints on the late time expansion rate}\label{section3}

Let us now confront the background expansion \eqref{Hsquare} sourced by the numerical solution of equation \eqref{Omegadot} which is subjected to the constraint \eqref{Omega0} against available observational data. Our analysis will be similar to Ref. \cite{Velten:2017hhf} but now adding Galactic globular clusters age constraints and the galaxy cluster gas mass fraction bounds on the model free parameters. Anticipating one of our results, such new information will be very important to revisit the main conclusion of Ref. \cite{Velten:2017hhf}. 

We will consider two different $f(R,T)$ models:
\begin{itemize}
    \item $f(R,T)=R+ \gamma_n T^{n}$;
    \item $f(R,T)= R + \alpha e^{\beta T}$.
\end{itemize}
Each model has three free parameters, one more than the flat $\Lambda$CDM model. The background expansion of the latter is described in terms of $H_0$ and $\Omega_{0}$. The cosmological constant fractionary density is $\Omega_{\Lambda}= 1-\Omega_0$. In the modified gravity scenarios studied here, the quantity $\Omega_{0}$ is replaced by a combination of $\alpha\, (\gamma_n)$ and $\beta \, (n)$ according to \eqref{Omega0}.

Our goal is to find a concordance region in the free parameter space for each model. For this task we shall use three different observational information.

{\it Age of the universe:} For a given expansion rate $H$, the age of the universe $t_U$ is calculated via integration 
\begin{equation}
    t_U=\int^1_0\frac{d\tilde{a} }{\tilde{a} H(\tilde{a})},
\end{equation}
where today's scale factor has been set $a_0=1$. Age constraints can be used as a simple tool to discriminate between viable and non-viable cosmologies. In this work we will adopt a minimum and obvious requirement that $t_U$ can not be smaller than the estimated age of astrophysical objects. Of course, the universe can not be younger than the structures it contains. Recent age estimations of Galactic globular clusters have placed the bounds \cite{
Valcin:2021jcg,Valcin:2020vav}
\begin{equation}
t_{glob}=13.5^{+0.16}_{-0.14} \,(stat.) \pm 0.5 \,(sys.) .
\end{equation}
We will use the above bounds to exclude modified gravity parameters yielding to young universes.    

{\it Gas fraction in galaxy clusters:} The {\it Chandra} measurements of X-ray from galaxy clusters is a powerful tool to constrain the temperature, gas density
and mass profiles of galaxy clusters. Such quantities are sensitive to the amount of the gas mass fraction in such systems and can be linked to the cosmological baryon to total matter ratio $\Omega_{b0}/{\Omega_{0}}$ \cite{Mantz:2014xba}. By relying on these bounds and associating the total matter to the parameter $\Omega_0$ appearing in (\ref{Omega0}), we can indirectly constrain the modified gravity parameters. Then, from the results presented in Ref. \cite{Mantz:2014xba} we shall demand
\begin{equation}\label{globular}
    0.23 < \Omega_0 < 0.31.
\end{equation}

{\it Cosmic Chronometers:} A widely used technique to measure the non-local expansion rate of the universe is to obtain the differential age of certain galaxies via the age of their stellar population. This method has allowed us to obtain measurements for $H(z)$ reaching up to redshifts around $z\sim 2$. This method, proposed in \cite{Jimenez:2001gg}, can be understood by the relation
\begin{equation}
    H(z)=-(1+z)\frac{dz}{dt},
\end{equation}
providing $H(z)$ at some redshift $z$ via the relation between differential cosmic ages of objects $dt$ within certain differential redshift range $dz$.

\begin{figure*}[t]
\includegraphics[scale=.33]{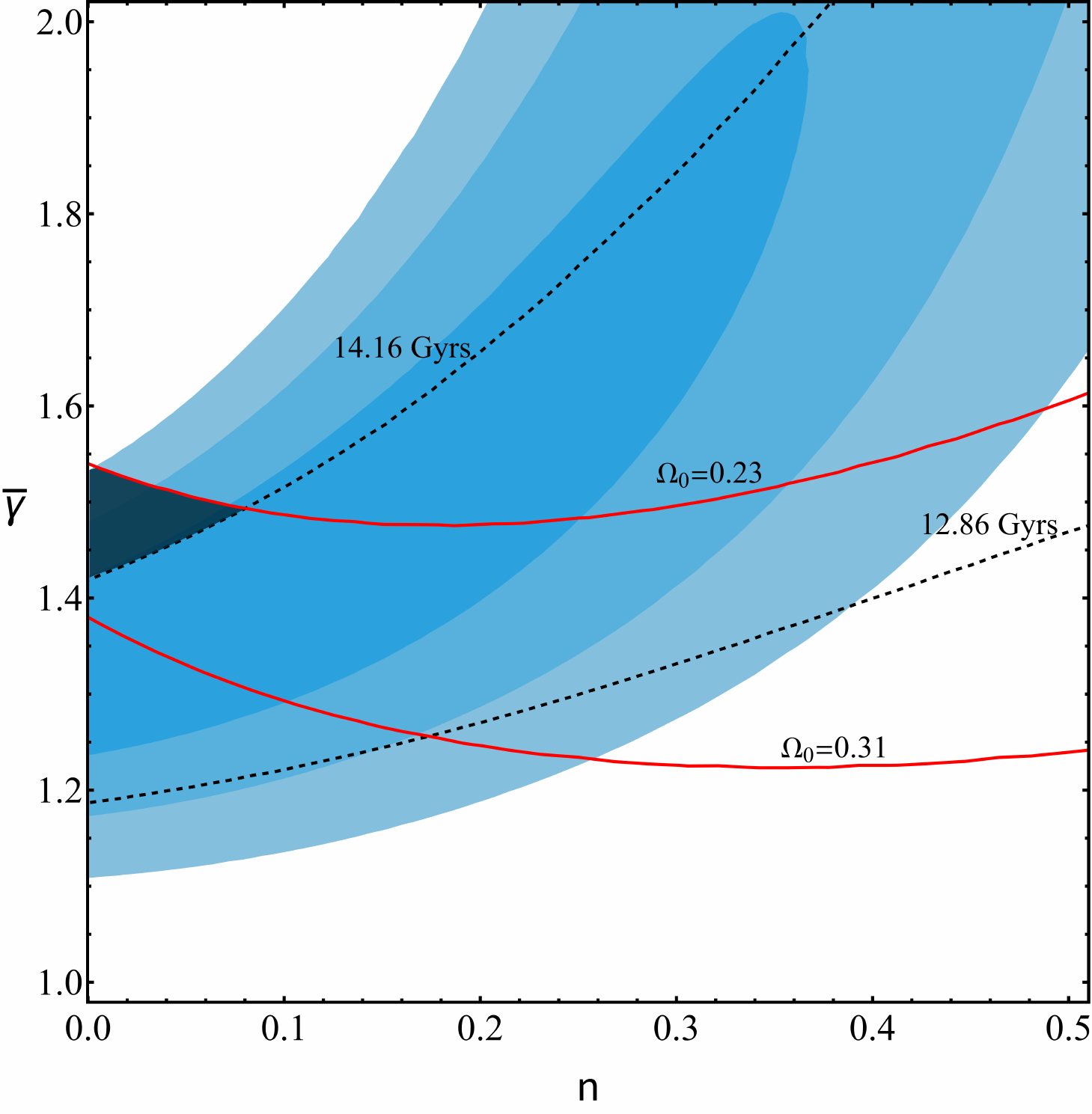}
\includegraphics[scale=.352]{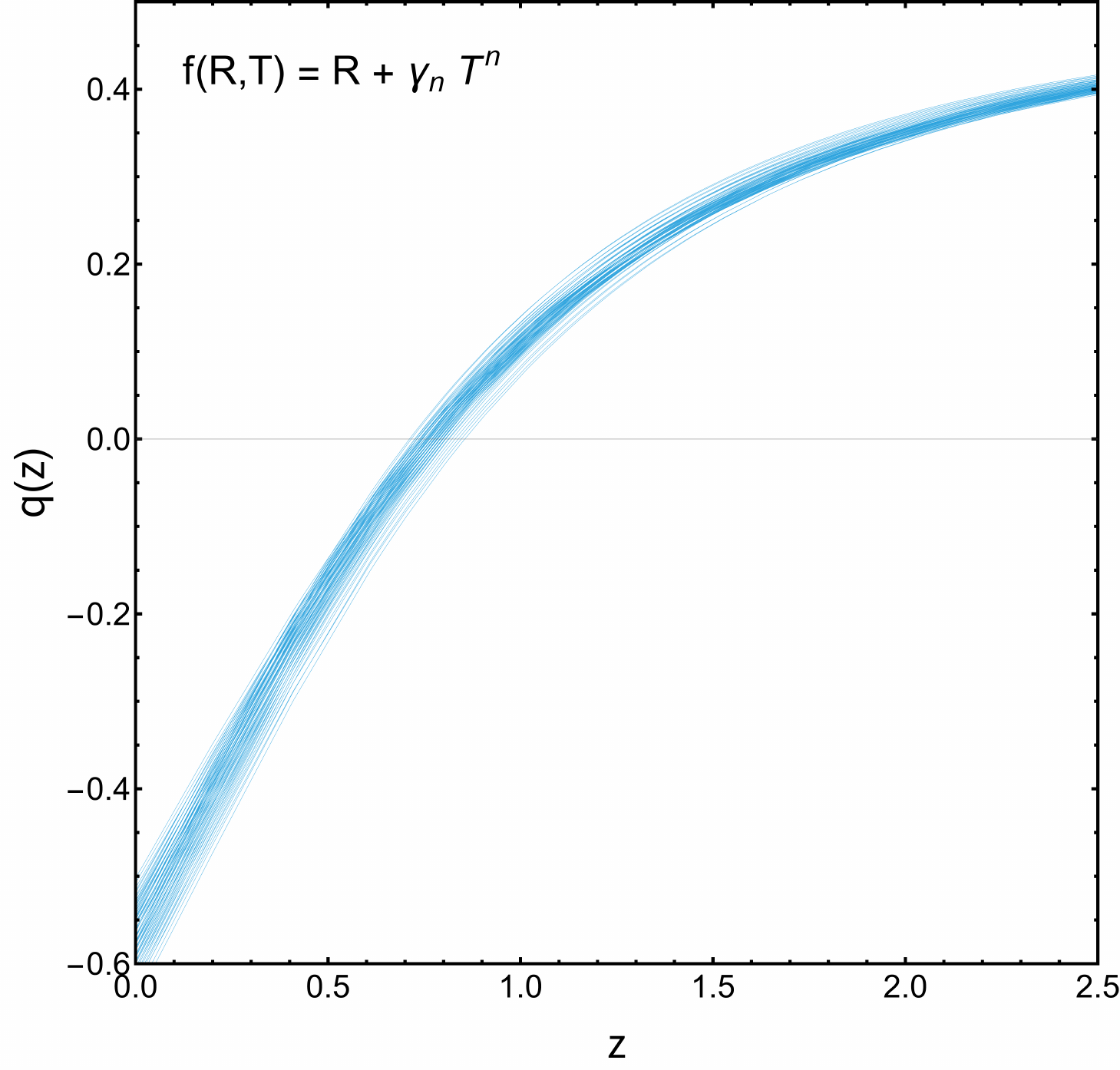}
\caption{Constraints on the free parameters of the power law model $f(R,T)= R + \gamma_{n} T^{n}$. In the left panel, the blue contours show $1\sigma$, $2\sigma$ and $3\sigma$ regions of statistical confidence level. Dashed lines represent the parameter values for which the universe is $12.86$ and $14.16$ Gyrs old. Red lines are maximum and minimum bounds on $\Omega_0$. The darker blue region represents the parameter space concordance region. In the right panel we plot the deceleration parameter as a function of the redshift $q(z)$ for sets of $\left\{ \bar{\gamma}_n, n \right\}$ values within the  darker blue region of the left panel.}
    \label{fig1}
\end{figure*}

We shall use in our analysis the data set available in Table 1 of Ref. \cite{Vagnozzi:2020dfn} in order to calculate statistical confidence contours for the modified gravity background dynamics studied in the last section.

We show in the left panel of Fig. \ref{fig1} the free parameter space for the power law model. In our analysis we will fix $H_0=67.4$ km s$^{-1}$ Mpc$^{-1}$ \cite{Planck:2018vyg}. We have checked that changing $H_0$ values around it yields a mild impact in our final conclusions. In this left panel dashed lines show age contours of $12.86$ Gyrs and $14.16$ Gyrs, red lines show $\Omega_0$ contours fixing the limits as in (\ref{Omega0}) and blue regions display the $1\sigma$, $2\sigma$ and $3\sigma$ statistical confidence level contours resulting from the likelihood function obtained from the $H(z)$ data. 

 To stay on the conservative side, let us consider that the universe should be older than $14.16$ Gyrs. This excludes a large region of the parameter space. The crossing of all such information i.e., an universe older than $14.16$ Gyrs, the parameter $\Omega_0$ within the bounds given by (\ref{globular}) and inside at least the $3\sigma$ region provides a narrow accepted parameter space value given by the darker blue region in this figure.

\begin{figure*}[t]
\includegraphics[scale=.328]{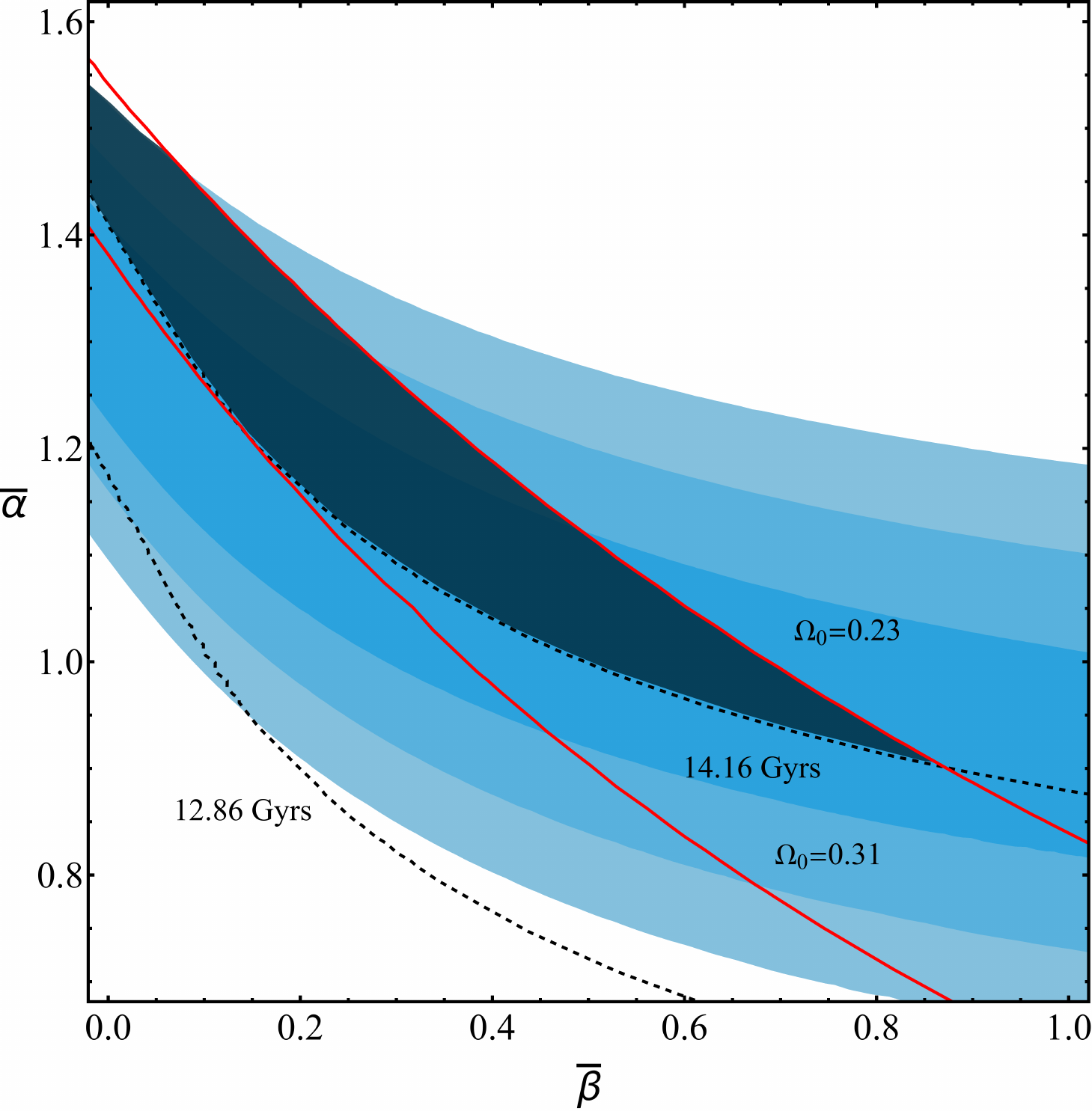}
\includegraphics[scale=.35]{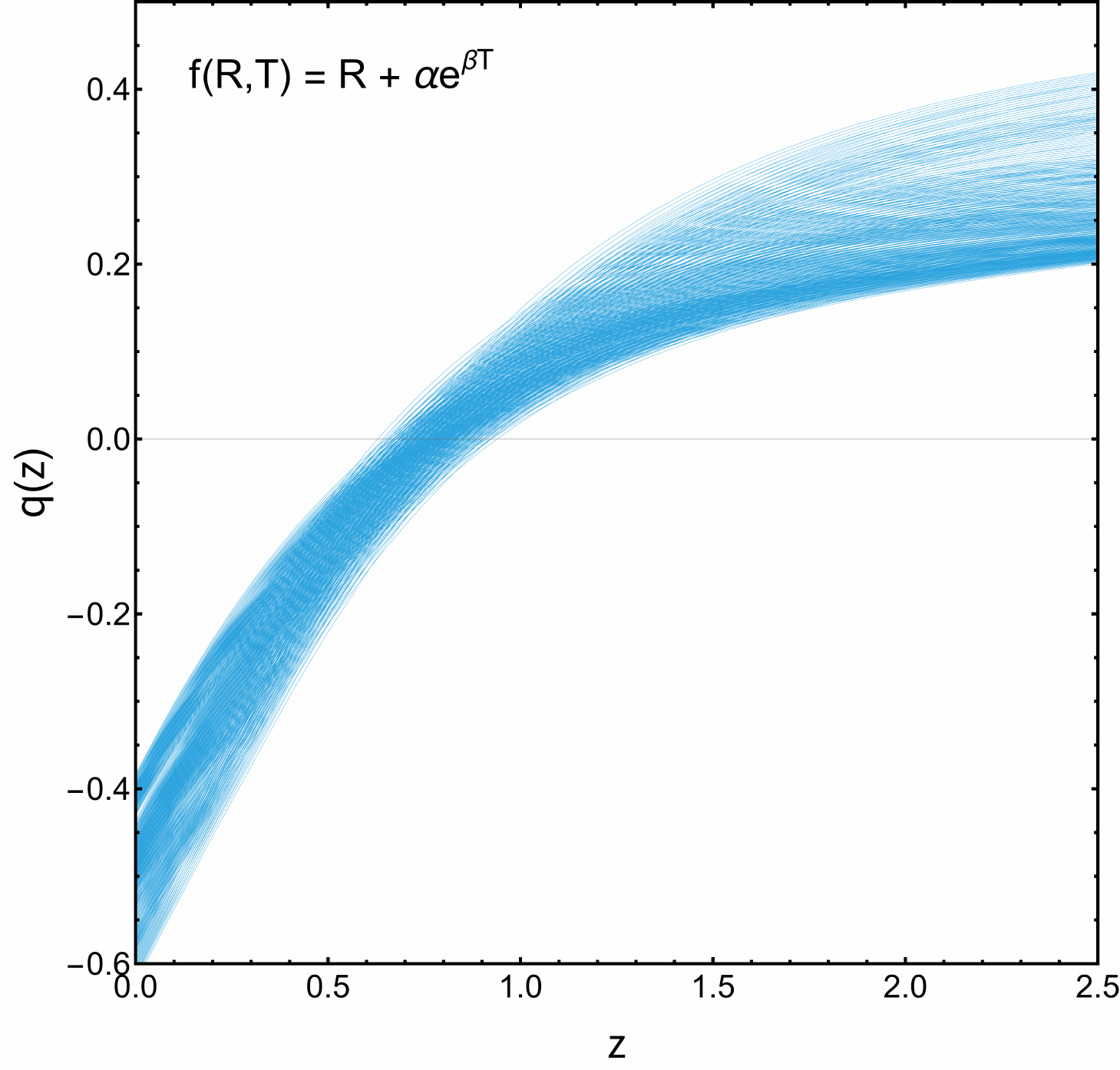}
    \caption{Constraints on the free parameters of the exponential model $f(R,T)= R + \alpha e^{\beta T}$. In the left panel, the blue contours show $1\sigma$, $2\sigma$ and $3\sigma$ regions of statistical confidence level. Dashed lines represent the parameter values for which the universe is $12.86$ and $14.16$ Gyrs old. Red lines are maximum and minimum bounds on $\Omega _{0}$. The darker blue region represents the parameter space concordance region. In the right panel we plot the deceleration parameter as a function of the redshift $q(z)$ for sets of $\left\{\bar{\alpha}, \bar{\beta} \right\}$ values within the  darker blue region of the left panel.}
    \label{fig2}
\end{figure*}

For such concordance range of parameters values found in the darker blue region, one can also verify how the expansion rate transits from the decelerated phase to the accelerated one via the definition of the deceleration parameter
\begin{equation}
    q(z)= -1 -\frac{\dot{H}}{H^2}.
\end{equation}
We then plot in the right panel in Figs. \ref{fig1} the deceleration parameter $q(z)$ as a function of the redshift $z$. This figure shows a collection of tiny blue curves computed using allowed parameter values found in the concordance darker blue region of the left panel.

Fig. \ref{fig2} shows results for the exponential model using the same structure as described in Fig. \ref{fig1}.
 
 In the case where either $n$ or $\beta$ vanish, $\gamma_n$ and $\alpha$ play the rôle of a cosmological constant in the gravitational action, respectively. One can associate $\bar{\gamma}_n$ and $\bar{\alpha}$ values to twice the cosmological constant fractionary parameter i.e., one can expect preferred values around $\bar{\gamma}_n \sim \bar{\alpha} \sim 2\Omega_{\Lambda}\sim 1.5$. In this limiting case the observational allowed region in both figures agree with this estimation.

Then, as one can see in both figures, $f(R,T)$ cosmologies have a viable parameter space to describe the late time cosmological observables.

\section{Adding radiation}\label{section5}

In this section we discuss the impact of adding radiation to the minimally-coupled model analysed in section \ref{section3}. Since radiation is a traceless fluid, one could in principle expect that there will be no impact at all on the cosmological dynamics. This is indeed the case for the early universe where the effective energy density is well approximated by a true radiative fluid. Therefore, any investigation about the impact of $f(R,T)$ models on the early universe features is soundless.  However, let us now investigate in more details how the entire cosmological background dynamics will involve in $f(R,T)$ gravity when both matter and radiation compose the total energy density tensor.

It is worth noting that equations deduced in section \ref{section2} refers to the total energy density $\rho$. Our task now is to decomposed it as the sum of radiation and matter i.e.,
\begin{equation}
    \rho=\rho_m+\rho_r.
\end{equation}
The equations of state of both fluids will be respectively $p_m=0$ and $p_r=\rho_r/3$. Also, the effective equation of state parameter $\omega$ entering the background relations \eqref{Hsquare} and \eqref{eqf4} shall be replaced by
\begin{equation}
    \omega=\frac{p_m+p_r}{\rho_m+\rho_r}=\frac{\rho_r}{3(\rho_m+\rho_r)}.
\end{equation}
Therefore, in presence of radiation and matter fields the final background relations are
\begin{equation}
    \frac{H^{2}}{H_{0}^{2}} = \left(\Omega _{m} + \Omega _{r}\right)\left(1 + f _{T}/\kappa ^{2}\right) + \frac{f_{T}}{3 \kappa ^{2}}\Omega _{r} + \frac{f(T)}{2\kappa ^{2}\rho _{0}},
\end{equation}
and
\begin{equation}
    \dot{\Omega}_{m}\left[1 + \frac{1}{\left(1 + f_{T}/\kappa ^{2}\right)}\left(\frac{f_{TT}\rho _{0}\Omega _{m}}{\kappa ^{2}} + \frac{f_{T}}{2\kappa ^{2}}\right)\right] + 3H\Omega _{m} + \dot{\Omega}_{r}\left[1 + \frac{1}{3\left(1 + f_{T}/\kappa ^{2}\right)}\left(\frac{4f_{TT}\rho _{0}\Omega _{r}\dot{\Omega}_{m}}{\kappa ^{2}\dot{\Omega}_{r}} + \frac{f_{T}}{\kappa ^{2}}\right)\right] + 4H\Omega _{r} = 0.
\end{equation}
The above equation represents the total energy conservation. In order to solve it let us split it into two different equations and demand that both should be satisfied simultaneously. The same procedure applies in the standard cosmological model. Hence the resulting set of equations is   
\begin{eqnarray}\label{consomegarad}
    \dot{\Omega}_{m}\left[1 + \frac{1}{\left(1 + f_{T}/\kappa ^{2}\right)}\left(\frac{f_{TT}\rho _{0}\Omega _{m}}{\kappa ^{2}} + \frac{f_{T}}{2\kappa ^{2}}\right)\right] + 3H\Omega _{m} =0 \\ 
    \label{consradomega}
     \dot{\Omega}_{r}\left[1 + \frac{1}{3\left(1 + f_{T}/\kappa ^{2}\right)}\left(\frac{4f_{TT}\rho _{0}\Omega _{r}\dot{\Omega}_{m}}{\kappa ^{2}\dot{\Omega}_{r}} + \frac{f_{T}}{\kappa ^{2}}\right)\right] + 4H\Omega _{r} = 0.
\end{eqnarray}

In the GR limit both equations reduce to the usual conservation laws leading to solutions $\Omega_{m}\propto a^{-3}$ and $\Omega_{r}\propto a^{-4}$ as expected. 

Note also that since $T = \rho _{m}$, in the family of $f(R,T)$ theories the $\Omega_m$ evolution does not depend on $\Omega_r$. Given the modified gravity parameters one can solve \eqref{consomegarad} to find $\Omega_m$. In the present case this will be performed numerically. However, in $f(R,T)$ cosmologies, the radiation density, obeying to  \eqref{consradomega}, will necessarily interact with the matter component. There is a source term proportional to $\dot{\Omega}_m$. This implies that the radiation component no longer scales according to the $a^{-4}$ law.

\begin{figure*}[t]
\includegraphics[scale=.45]{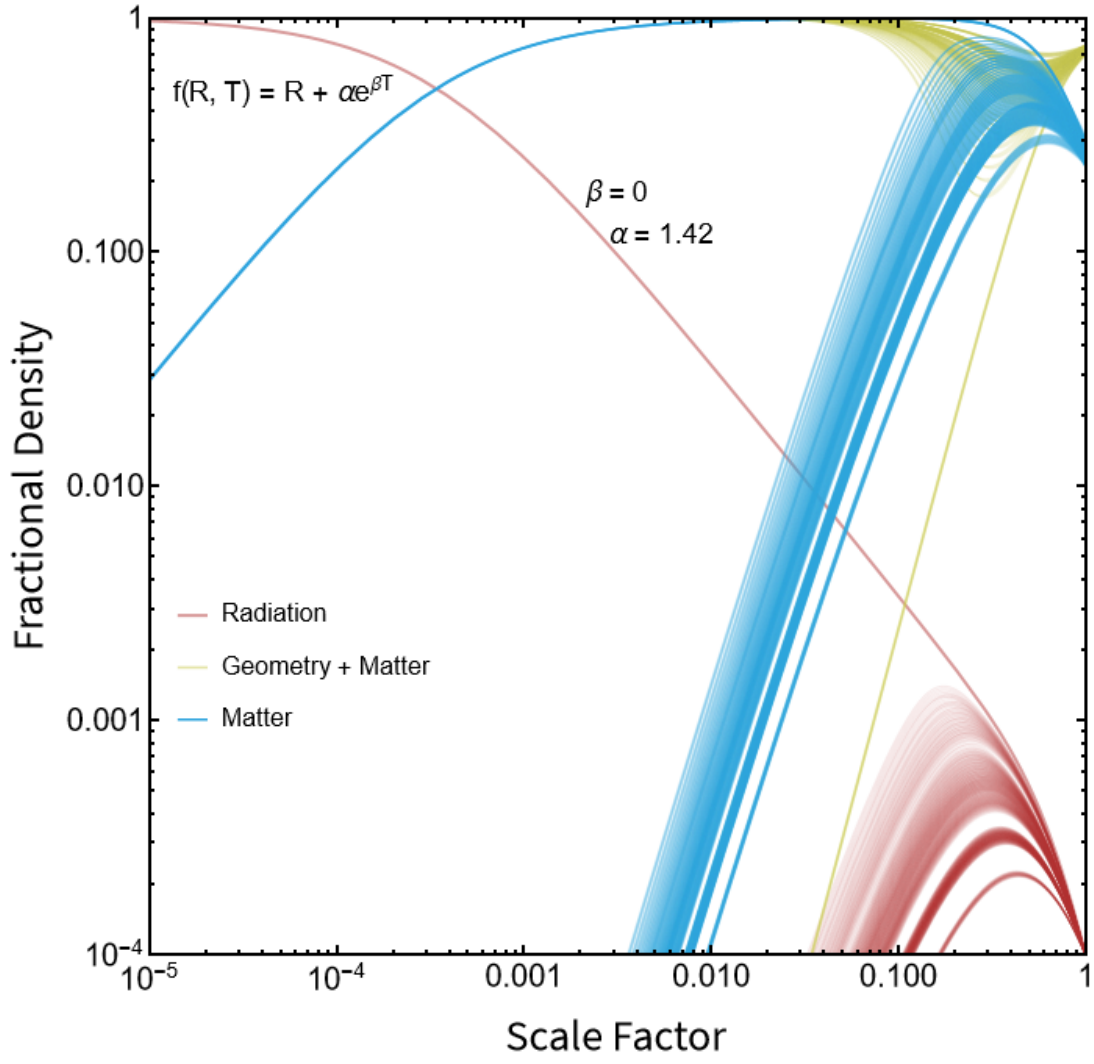}
\includegraphics[scale=.45]{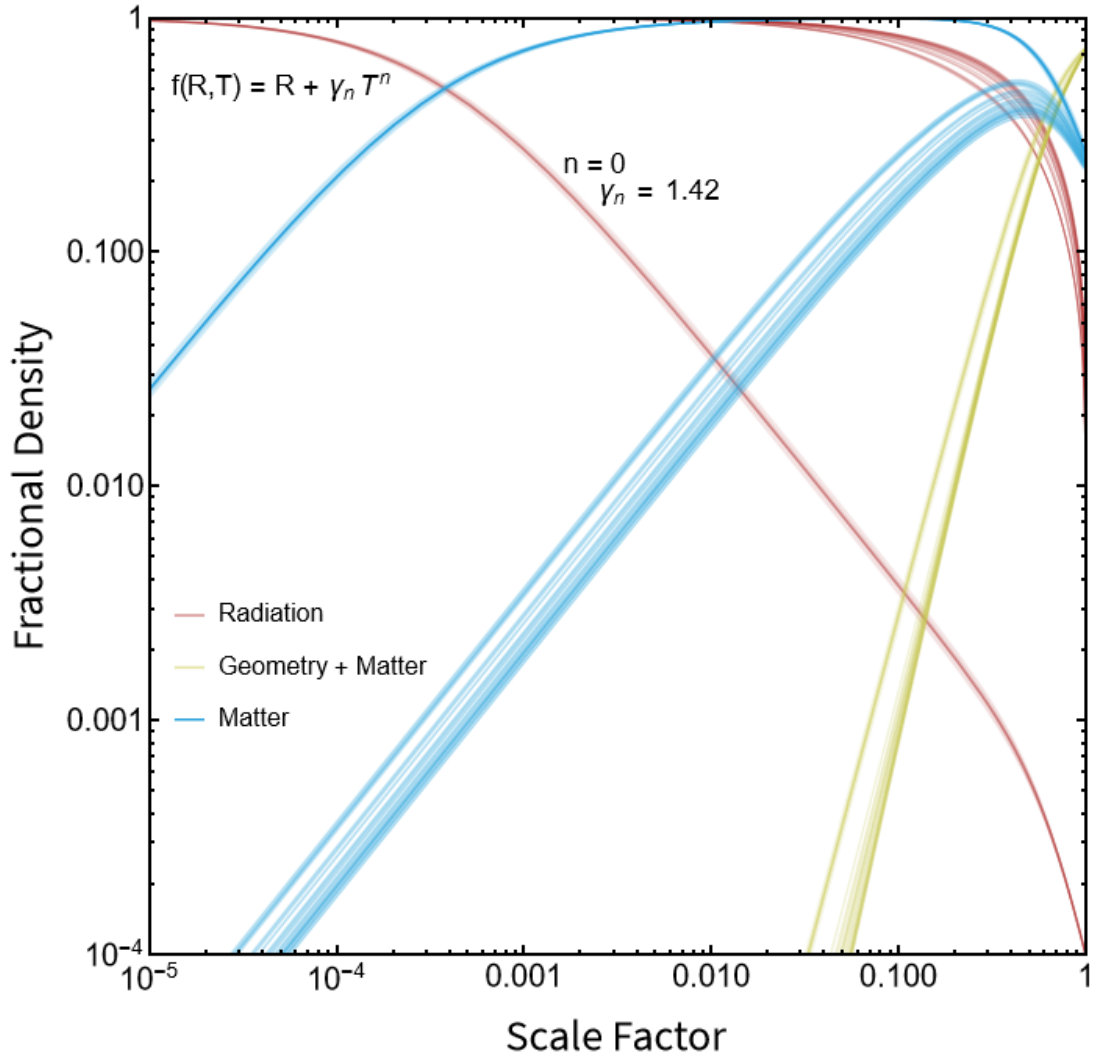}
    \caption{Evolution of the fractionary energy densities as a function of the scale factor. }
    \label{fig3}
\end{figure*}

We show in \ref{fig3} the evolution of the fractionary energy densities for the different cosmic components as a function of the scale factor. Red (blue) curves represent the radiation (matter) evolution. The yellow curves show how the effective energy density associated to the geometric term \eqref{rhobar} contribution evolves. Left (right) panel considers the exponential (polynomial) model. For all curves the present values of the radiation energy density has been fixed as $\Omega_r0=9.847 \times 10^{-5}$ according to Planck 2018 result. In both panels we plot the $\Lambda$CDM limit in the curve with denoted with the label $\beta=0, \alpha=1.42$ (left panel) and $n=0, \gamma_n=1.42$ (right panel). In the remaining curves we solve the dynamics given by \eqref{consomegarad} and \eqref{consradomega} for the same set of modified gravity parameters values used to plot Figs. \ref{fig1} and \ref{fig2} i.e., the parameter values filling the dark blue region in theses figures. All cases with $\beta\neq 0$ and $n \neq 0$ are physically meaningless. In common, for $\beta\neq 0$ and $n \neq 0$, both models present a transition from the radiation dominated to the matter phase quite recently.  Though the generalized second law of thermodynamics is generally valid, independently of the specific interaction form between matter and radiation as shown in Ref. \cite{Jamil:2009eb}, this clearly impedes $\beta\neq 0$ and $n \neq 0$ models to explain, for instance, the cosmic microwave background acoustic peak structure and the growth of matter overdensities.

\section{The non-minimally coupled case: $f(R,T)= f_{1}(R) + f_1(R)f_2(T)$}\label{section4}

The cosmological background evolution in non-minimally coupled cases of the form
\begin{equation}
    f(R,T)= f_{1}(R) + f_1(R)f_2(T),
\end{equation}
has been investigated in Refs. \cite{Moraes:2017zgm,Sofuoglu:2022eut}. As shown in these references, the resulting modified Friedmann equations are such that the squared expansion rate depends on its derivative. Therefore, a constraining relation like \eqref{Omega0} can not be imposed to find out the $\Omega_0$ value. In references  \cite{Moraes:2017zgm,Sofuoglu:2022eut}, rather than solving the background dynamics numerically, analytical solutions are found by imposing that the expansion rate has a power law dependence on the cosmic time. 

Let us then try another approach for solving the background dynamics in non-minimally coupled $f(R,T)$ models. We explore the dynamical equations in the metric-affine formalism as firstly studied in Ref. \cite{Barrientos:2018cnx}.

However, in the metric-affine (or Palatini) formalism, the variation of the Ricci tensor is performed in terms of the connection, which means that the operator $\Delta _{\mu\nu}$ in equation \eqref{eq8} does not exist \cite{Barrientos:2018cnx}. With respect to the modified Friedmann equations, this result implies that the first and second time derivatives, $\dot{f}_{R}$ and $\ddot{f}_{R}$, disappear. 

In order to provide an explicit example, let us now show the background dynamics of the non-minimally coupled model by considering  
\begin{equation}
f(R,T)= f_{1}(R) + f_{1}(R)f_{2}(T) = \epsilon R + \lambda _{m}R T^{m}.
\end{equation}

The parameter $\epsilon$ allows one either to keep the Einstein-Hilbert term intact or to switch it off with a vanishing $\epsilon$.

From the above one finds the following cosmological dynamical equations in the metric-affine formalism 
\onecolumngrid
\begin{equation}\label{htn}
    3H^{2} \left[\epsilon + \lambda _{m}\rho ^{m}(1 - 3\omega)^{m} + 4\lambda _{m}m\rho ^{m}(1 - 3\omega)^{m -1}(1 + \omega)\right] = \kappa ^{2}\rho - 6\lambda _{m}m\rho ^{m}(1 - 3\omega)^{m -1}(1 + \omega)\dot{H},
\end{equation}
and
\begin{eqnarray}\label{hen}
  -2\dot{H}\left[\epsilon + \lambda _{m}\rho ^{m}(1 - 3\omega)^{m}\right] = \kappa ^{2}\rho \omega + 3H^{2}\left[\epsilon + \lambda _{m}\rho ^{m}(1 - 3\omega)^{m}\right].
\end{eqnarray}

Both expressions \eqref{htn} and \eqref{hen} are different from previous results presented in the literature based on the metric formalism \cite{Sofuoglu:2022eut,Moraes:2017zgm}. By combining \eqref{htn} and \eqref{hen} we find the background expansion rate in in non-minimally coupled $f(R,T)$ based on the metric-affine formalism obeying to
\begin{eqnarray}
    \frac{H^{2}}{H^{2}_{0}} &=& \left[\epsilon + \tilde{\lambda}_{m}\Omega ^{m}(1 - 3\omega)^{m} + \tilde{\lambda}_{m}m\Omega ^{m}(1 + \omega)(1 - 3\omega) ^{m-1}\right]^{-1}\left\{\Omega + \frac{6\tilde{\lambda}_{m}m\Omega(1 + \omega)(1-3\omega)^{m -1}\omega \Omega ^{m}}{2[\epsilon + \tilde{\lambda}_{m}\Omega ^{m} (1 - 3\omega)^{m}]}\right\}.
\end{eqnarray}
The dimensionless parameter appearing above has been defined as 

\begin{equation}
    \tilde{\lambda}_{m} = \lambda _ {m}\rho ^{m}_{0}.
\end{equation}
It is worth noting that the limiting case $m=0$ is not equivalent to the $\Lambda$CDM cosmology. Instead, this case means a simple redefinition of the gravitational coupling $\kappa^2$ by a constant value $(\epsilon+\tilde{\lambda}_{0})$.

Contrarily to the metric formalism, the above equation for $H$ obtained in the metric-affine approach allows one to set the constraining relation between $\Omega_0$ and the modified gravity parameters as in (\ref{Omega0}) i.e.
\begin{eqnarray}\label{Omega0coupled}
    1=  \left[\epsilon + \tilde{\lambda}_{m}\Omega_0 ^{m}(1 - 3\omega)^{m} + \tilde{\lambda}_{m}m\Omega_0 ^{m}(1 + \omega)(1 - 3\omega) ^{m-1}\right]^{-1}\left\{\Omega_0 + \frac{6\tilde{\lambda}_{m}m\Omega_0(1 + \omega)(1-3\omega)^{m -1}\omega \Omega_0 ^{m}}{2[\epsilon + \tilde{\lambda}_{m}\Omega_0 ^{m} (1 - 3\omega)^{m}]}\right\}.
\end{eqnarray}

In order to obtain the dynamical evolution for the matter density parameter $\Omega$ one solves its conservation law expressed by the equation
   \begin{equation}\label{equationnonminimally}
    \dot{\Omega} + 3H\Omega (1 + \omega) = \left\{ 1 + \frac{\chi\omega}{\zeta \left[\epsilon + \tilde{\lambda} _ {m}\Omega ^{m}(1 - 3\omega)^{m}\right]}\right\}^{-1}\left\{(1 + \omega)\left(\frac{\dot{\chi}}{\chi} - \frac{\dot{\zeta}}{\zeta}\right)\Omega - \dot{\Omega} \omega + \frac{\dot{\Omega}\chi \omega}{\zeta \left[\epsilon + \tilde{\lambda} _ {m}\Omega ^{m}(1 - 3\omega)^{m}\right]}\right\},
\end{equation}
where we have defined
\begin{equation}\label{chi}
\chi = \left[\epsilon + \tilde{\lambda} _{m}\Omega ^{m}(1 - 3\omega)^{m} + \tilde{\lambda} _{m}m\Omega^{m}(1 - 3\omega)^{m -1}(1 + \omega)\right],
    \end{equation}
    and
   \begin{equation}\label{zeta}
        \zeta = 1 + \frac{3\tilde{\lambda} _{m}(1 + \omega)(1 -3\omega)^{m - 1}\omega\Omega ^{m}}{\epsilon + \tilde{\lambda} _{m}\Omega ^{m}(1 - 3\omega)^{m}}.
   \end{equation}

 Now, by assuming a pressureless matter component, $\omega = 0$, the temporal derivative of \eqref{chi} and \eqref{zeta} reduces to, respectively
$\dot{\chi} = \tilde{\lambda}_{m}m(m + 1)\Omega ^{m - 1}\dot{\Omega}
$ and $      \dot{\zeta} = 0 $.  Thus, using these results the equation \eqref{equationnonminimally} becomes
\begin{equation}\label{contcoupled}
    \dot{\Omega} + 3H\Omega = \frac{\tilde{\lambda}_{m}m(m + 1)\dot{\Omega}\Omega ^{m}}{\epsilon + \tilde{\lambda}_{m}\Omega ^{m}(m + 1)}.
\end{equation}

The limiting cases leading to conservative models are easily identified i.e., either $\tilde{\lambda}_m=0$ or $m=0, -1$. 
Once again, using the same strategy as in the previous section, we can apply to the bounds provided in (\ref{Omega0}) to the constraining relation (\ref{Omega0coupled}) to find the allowed modified gravity parameters values. We then solve numerically equation \eqref{contcoupled} for the allowed modified gravity parameter values and find out that all resulting cosmological dynamics are inconsistent with data. In particular, all cases have  pure Einstein-de Sitter like evolution for all redshifts given no transition to an accelerated epoch.

\section{Conclusions}\label{section6}

Our goal in this work is to revisit the cosmological background expansion in $f(R,T)$ theories of gravity focusing on the minimally coupled model $f(R,T)=f_1(R)+f_2(T)$. The latter case with $f_1(R)=R$ represents a particular situation within modified gravity theories in which the expansion rate $H$ can be written explicitly in terms of the total matter density parameter $\Omega$ as well as the modified gravity free model parameters. This allows us to find the constraining relation \eqref{Omega0} which is the most important relation in this work. It is worth mentioning that generic $f(R)$ theories $H$ depends on $\dot{\Omega}$ and therefore such constraining relation is absent.

The validity of \eqref{Omega0} therefore allows one to place stringent bounds on the model free parameters. By interpreting $\Omega_0$ as the effective fractionary matter density parameter, and demanding it is bounded by the gas mass fraction estimations in galaxy clusters given by (\ref{globular}) one can directly constrain the modified gravity parameters appearing in (\ref{model}). An additional requirement concerns the age of the universe. The modified gravity parameters should provide an age for the universe larger than the estimated age of Galactic globular clusters.  Since a $4$ free modified gravity parameters is not competitive, we have studied $2$ free parameters models namely, the polynomial and the exponential ones. Figs. \ref{fig1} and \ref{fig2} summarize our main findings. There is indeed a tiny modified gravity parameter space allowed by background cosmological data. It is important to stress that this conclusion is opposite to the one presented in  \cite{Velten:2017hhf}, by one of the authors of the present work, which ruled out the power law class of $f(R,T)$ theories. The conclusion in \cite{Velten:2017hhf} was based on a qualitative analysis of low redshift data, and the high redshift behavior of $f(R,T)$ models was extrapolated without proper comparison to actual observational data. Therefore, it can be said that the analysis in \cite{Velten:2017hhf} was incomplete in some sense. In the current work, we have adopted a better statistical procedure and quantitatively accounted age constraints that played a fundamental role in this analysis. In summary, compared to Ref. \cite{Velten:2017hhf}, we have provided a more sophisticated analysis, which indicates that there is indeed a region in the free parameter space around $n=0$ or $\beta=0$ that is allowed by observational data. Of course, in order to determine the viability of $f(R,T)$ cosmologies, our next step is to analyze scalar perturbations and conduct a proper comparison with available large scale structure data. This will be the subject of a future investigation. 

The analysis discussed above only concerns the late time cosmological where usually one neglects the contribution from radiation. Indeed, we have chosen to exclude it from the background dynamics since the radiation-dominated epoch lasted only approximately $50,000$ years. The radiation fluid has minimal impact on the overall age of the universe and on the late time cosmological observables. However, a full model should include the radiation component. We discuss again below the implications of the latter scenario for $f(R,T)$ gravity.

Before our final conclusion, let us comment on the non-minimally coupled case. The constraining relation (\ref{Omega0}) does not exist in the non-minimally models using the metric formalism since $H^2$ depends on the derivative $\dot{H}$. This is also the situation in $f(R)$ theories. However, in the metric-affine approach the modified Friedmann equations lead to the constraining relation (\ref{Omega0coupled}). The background dynamical evolution of the non-minimally coupled case, on the other hand, are non viable since all modified gravity parameter values allowed by the gas fraction bounds are consistent with Einstein-de Sitter cosmologies for all redshifts i.e., the universe does not transits to an accelerated phase as supported by current cosmological observables.

Now, let us focus on the role played by the inclusion of the radiation component. The main conclusion of this work is encoded in section \ref{section5} where we have incorporate the radiation component in our analysis of the background dynamics in $f(R,T)$ theories. We have shown that, by interpreting the total energy density $\rho$ as the sum of matter and radiation energies, necessarily, the background dynamics turns into a interacting cosmological model between matter and radiation. Hence, the energy density of the radiation component does not scales according to the standard law $a^{-4}$. Minimal contributions from the trace dependent part $f_2(T)$ lead to inconsistent cosmological scenarios as shown and discussed in the panels of Fig. \ref{fig3}. 

Finally, putting together the findings of this work with the recently published results on the behavior of $f(R,T)$ theories in the solar system \cite{Bertini:2023pmp}, we conclude that $f(R,T)$ theories should be ruled out.

\begin{acknowledgments}

The authors thank FAPEMIG/FAPES/CNPq and CAPES for financial support. We thank Rodrigo von Marttens and Jailson Alcaniz for useful correspondence. The authors also thank the anonymous referee for pointing out the relevance role played by the radiation component. 
\end{acknowledgments}

\end{document}